\documentclass{IEEEtran}
\usepackage{cite}
\usepackage{amsmath,amssymb,amsfonts}
\usepackage{graphicx}
\usepackage{textcomp,nicefrac}
\usepackage{amssymb}

\usepackage{lineno,hyperref}
\modulolinenumbers[5]
\usepackage{adjustbox}
\usepackage{graphicx}
\usepackage{dcolumn}
\usepackage{bm}
\usepackage{rotating}
\usepackage{xcolor}

\usepackage{tikz}
\usetikzlibrary{positioning}
\usetikzlibrary{shapes,arrows,arrows,positioning,fit}
\usepackage{graphicx}
\usepackage{chngcntr}
\usepackage{mathtools}
\usepackage{amsfonts}

\usepackage{subfig}
\DeclareUnicodeCharacter{03B3}{$\gamma$}
\usepackage{textgreek}

\def\BibTeX{{\rm B\kern-.05em{\sc i\kern-.025em b}\kern-.08em
T\kern-.1667em\lower.7ex\hbox{E}\kern-.125emX}}
\begin{document}
\title{Development of Indigenous Pulse-Shape Discrimination Algorithm for Organic Scintillation detectors.}

\author{Annesha Karmakar, G. Anil Kumar* \IEEEmembership{Senior Member, IEEE}, Bhavika, V. Anand, Anikesh Pal.
\thanks{Annesha Karmakar is with Nuclear Engineering and Technology Program, Department of Mechanical Engineering, Indian Institute of Technology-Kanpur, India in collaboration with Radiation Detectors and Spectroscopy Laboratory, Department of Physics, Indian Institute of Technology Roorkee, India. (e-mail: annesha@iitk.ac.in). G. Anil Kumar is with Radiation Detectors and Spectroscopy Laboratory, Department of Physics, Indian Institute of Technology Roorkee, India. (e-mail: anil.gourishetty@ph.iitr.ac.in ). Anikesh Pal is with Department of Mechanical Engineering, Indian Institute of Technology-Kanpur India. (e-mail: pala@iitk.ac.in).}}
\maketitle

\begin{abstract}
The use of programmable hardware devices is imperative for digital based pulse shape discrimination (PSD) to differentiate between various types of radiation. This work reports the development of a PSD algorithm based on tail area and total area, eliminating the need for programmable hardware. The pulses were collected using BC501 detector and Pu-Be source from a digitizer in the oscilloscope mode. The algorithm performs crucial functions such as pulse normalization, shaping, identification and removal of multiple peaks and threshold determination. The algorithm provides neutron and $\gamma$-ray counts, scatter plot, and FoM.  In order to test the efficacy of our proposed algorithm, pulses were collected from a different source-detector setup comprising BC501A detector and an Am-Be source from a digitizer in the oscilloscope mode and Charge Integration (CI) mode. The results obtained from our proposed algorithm and CI method clearly indicates  a good agreement in terms of number of neutrons and $\gamma$-rays and Figure-of-Merit (FoM), thus providing cost-effective alternative method for neutron and $\gamma$-ray discrimination, offering flexibility and accuracy without specialized hardware.
\end{abstract}


\section{Introduction}
Digital pulse shape discrimination (PSD) is a widely used technique in nuclear physics, radiation detection, and signal processing. It analyzes the shape or pattern of signals to discriminate between different types of radiation. Digital PSD examines waveform attributes such as rise time, fall time, and width to categorize pulses. This approach provides more reliable and precise recognition than analog methods, allowing for improved discrimination between radiation signals. Many investigations have been conducted on the subject of classical digital techniques of PSD.
The studies emphasized the use of programmable hardware modules for implementing these digital PSD techniques. Additionally, charge integration was highlighted as a commonly used method in PSD for differentiating radiation pulses, such as gamma rays and neutron pulses, based on their shapes and charge distributions. Several references support the use of digital PSD techniques with programmable modules for radiation discrimination \cite{Paweczak2013, Wang2014, Balmer2015, Nakhostin2019, Doucet2020}. 

In order to avoid the dependence on programmable hardware for PSD, the present work aims to develop a PSD algorithm based on tail area and total area of the pulse. This algorithm performs all the necessary functions namely, normalisation of discrete pulses, shaping and smoothing, identifying and removing multiple peaks in a given record length, determining the start of tail area and optimal threshold, displaying the number of neutrons and $\gamma$-rays along with the scatter plot and FoM. These pulses were collected using BC501 detector and Pu-Be source from a digitizer in the oscilloscope mode only. To assess algorithm efficacy, the algorithm was applied on a different BC501A detector and Am-Be source from digitizer in the oscilloscope mode. The discriminating capabilities and Figure-of-Merit (FoM) acquired from our proposed PSD algorithm are compared with the results obtained from the digitizer's built in CI method. 

\section{Experiments}
\subsection{Measurement for parent experiment }
The experimental setup involves a 5-Ci Pu-Be source positioned at a distance of 5 cm from the front face of a liquid scintillator BC501 (manufactured by Saint Gobain crystals) \cite{BC501} with a radius of 7.5 cm and a length of 12.5 cm. The source produces neutrons (up to 11 MeV), and $\gamma$-rays (up to 6 MeV) due to ($\alpha$,n) reactions \cite{LBNL, Knoll:1300754} as shown in equation \ref{eq:1} below.

    \begin{equation} \label{eq:1}
	    \alpha + ^{9}Be -> n+^{12}C + 5.7MeV  \
	\end{equation}	

To record the pulses of neutrons and $\gamma$-rays, the output of the photo-multiplier tube (PMT) of the detector is connected to a 1 GHz CEAN DT-5751 desktop digitizer \cite{DT5751}. A total of 57934 pulses were collected in the oscilloscope mode of the digitizer and saved in $.dat$ file format. This experiment which is labeled as `Parent Experiment' was conducted at the Radiation Measurement Laboratory, Indian  Institute of Technology Kanpur, India.
\subsection{Measurement for test experiment}
The experimental setup involves a 1.3 $\mu$Ci Cf-252 source was placed at a distance of 12 cm from a $3'' \times 3''$ BC501A detector. Discrete pulses were acquired for 45 mins from a 1 GHz, DT5751 digitizer in oscilloscope mode. A total number of 46323 pulses were collected and store in `.dat' format. Further, the digitizer calculated the PSD  based on the CI method giving us both the number of neutrons and $\gamma$-rays and also the FoM for the same pulses. This experiment, labeled as `Test experiment' and was conducted at Radiation Detectors and Spectroscopy Laboratory, Indian Institute of Technology Roorkee, India.
\section{Methodology of Tail-to-Total Area Algorithm }
PSD is important for accurately identifying radiation types, improving measurement accuracy, and enhancing detector performance. By leveraging the unique pulse shapes of different radiation types, PSD techniques contribute to a wide range of applications in radiation detection, measurement, and safety. 
The algorithm operates on discrete pulses obtained from a digital oscilloscope, which are stored in a `.dat file' for input. The algorithm's methodology is outlined through the following steps:
\subsection{Analysing discrete pulses}
\begin{figure}[ht]
	\centering
		\includegraphics[height =2.5in]{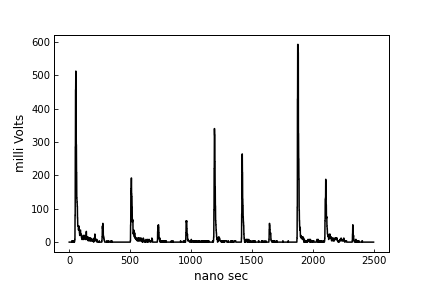}
		\caption{\label{fig:Pulses}Sample pulses observed in oscilloscope mode of digitizer}
	\end{figure}

The waveforms obtained from the digitizer are shown in figure \ref{fig:Pulses}. The pulses acquired from the digitizer are in form of data points (discrete pulse). These waveforms are similar in terms of rise time but are different in terms of amplitude and fall time. 
We identify the baseline of the pulses and subtract it from the data points of the waveforms to obtain normalised discrete pulses. Additionally clipped pulses are removed from the entire dataset.

\subsection{Shaping and smoothing the pulse}
\begin{figure}[ht]
 \centering
  \includegraphics[height = 2.5in]{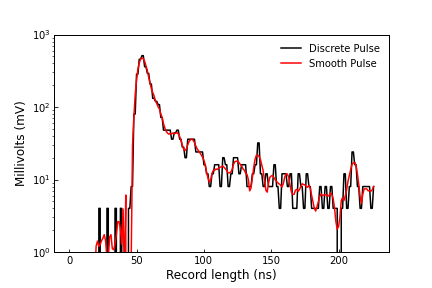}\\
 \caption{Superimposition of discrete and smoothed pulse to determine tail area}
 \label{fig:PulsesLOG}
 \end{figure}

The normalised pulses so obtained  have been shaped and smoothed using  the Savitzky-Golay function \cite{Savitzky1981} which can be represented mathematically by equation \ref{eq 2}. Here, B is the filter operator, f is the original spectrum with value $f[m]$, m = -2, -1, 0, 1, 2,.....  and b is the filter function and the brackets describe its discrete nature. As the pulses obtained are of finite nature, this function is the summation over finite intervals $|n|\leq N$.
	\begin{equation} \label{eq 2}
	    Bf[m] = \sum_{n=-\infty}^{\infty} b[n]f[m+n] \quad \textrm{for} \quad n\in N \
	\end{equation}
After smoothing, the discrete pulses and smoothed pulses are superimposed to determine the start of the tail area as shown in Figure \ref{fig:PulsesLOG} as an example for a single pulse.

\subsection{Pulse Pile-up handling}
To eliminate the pulse pile-up, the record length, also known as the window of each pulse, is verified using a loop for each pulse. The window is chosen as a typical practice based on observation in the oscilloscope mode of the digitizer, where the entire pulse is covered. The pulse was determined to be over-lapping or a pile-up when two or more peaks were observed within a particular window. To handle all overlapping pulses in the entire dataset, a loop was declared that ranged the entire length of the data set with an interval of window length and a conditional statement that said if the number of peaks was more than one, delete the data points from the entire dataset and move to the next window length; otherwise, consider the pulse.

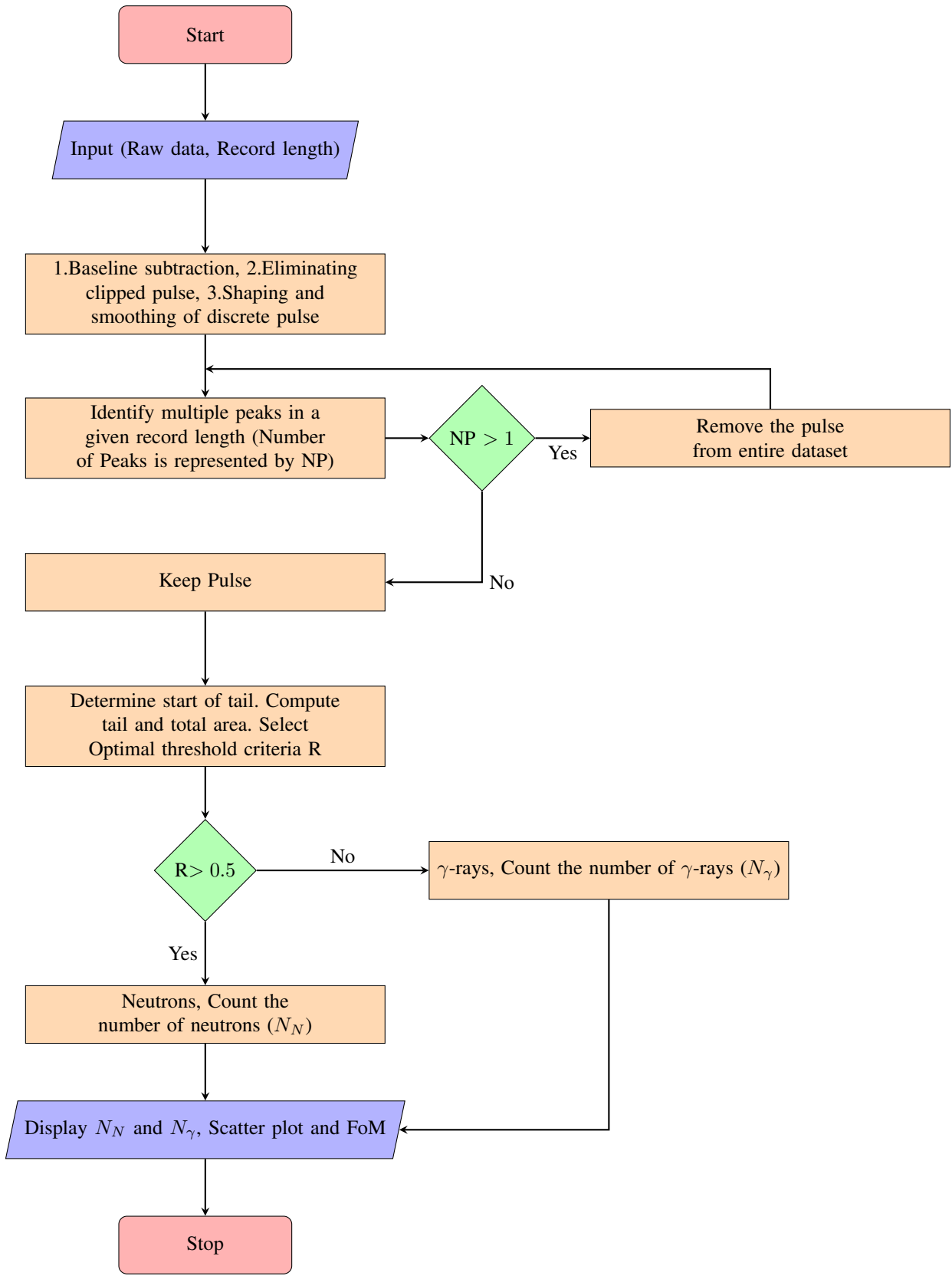
\begin{figure*}
\usetikzlibrary{shapes.geometric, arrows}
\usetikzlibrary{positioning}
\usetikzlibrary{shapes,arrows,arrows,positioning,fit}
\tikzstyle{startstop} = [rectangle, rounded corners, 
minimum width=3cm, 
minimum height=1cm,
text centered, 
draw=black, 
fill=red!30]

\tikzstyle{io} = [trapezium, 
trapezium stretches=true, 
trapezium left angle=70, 
trapezium right angle=110, 
minimum width=3cm, 
minimum height=1cm, text centered, 
draw=black, fill=blue!30]

\tikzstyle{process} = [rectangle, 
minimum width=1cm, 
minimum height=1cm, 
text centered, 
text width=6cm, 
draw=black, 
fill=orange!30]

\tikzstyle{decision} = [diamond, 
minimum width=1cm, 
minimum height=1cm, 
text centered, 
draw=black, 
fill=green!30]
\tikzstyle{arrow} = [thick,->,>=stealth]

\begin{tikzpicture}[node distance=2cm]

\node (start) [startstop] {Start};
\node (in1) [io, below of=start] {Input (Raw data, Record length)};
\node (pro1) [process, below of=in1,yshift=-0.5cm] {1.Baseline subtraction, 2.Eliminating clipped pulse, 3.Shaping and smoothing of discrete pulse};
\node (pro) [process, below of=pro1,yshift=-0.5cm] {Identify multiple peaks in a given record length (Number of Peaks is represented by NP)};
\node (dec1) [decision, right of=pro, xshift=2.8cm] {NP $>1$};
\node (proa) [process, right of=dec1, xshift=3cm] {Remove the pulse\\from entire dataset};
\node (prob) [process, below of=pro, yshift=-0.5cm] {Keep Pulse};
\node (pro2) [process, below of=prob, yshift=-0.5cm] {Determine start of tail. Compute tail and total area. Select Optimal threshold criteria R};

\node (dec2) [decision, below of=pro2, yshift=-0.5cm] {R$> 0.5$};

\node (pro3a) [process, below of=dec2, yshift=-0.5cm] {Neutrons, Count the number of neutrons ($N_N$)};

\node (pro3b) [process, right of=dec2, xshift=5cm] {$\gamma$-rays, Count the number of $\gamma$-rays ($N_\gamma$)};
\node (out1) [io, below of=pro3a] {Display $N_N$ and $N_\gamma$, Scatter plot and FoM };
\node (stop) [startstop, below of=out1] {Stop};

\node [coordinate, below of=pro1, node distance=1.3cm] (middle) {};

\draw [arrow] (start) -- (in1);
\draw [arrow] (in1) -- (pro1);
\draw [arrow]  (pro1) -- (pro);
\draw [arrow]  (pro)--(dec1);
\draw [arrow] (dec1) -- node[anchor=north]{Yes}(proa);
\draw [arrow] (dec1) |- node[anchor=west] {No} (prob);
\draw [arrow] (pro2) -- (dec2);
\draw [arrow] (dec2) -- node[anchor=east] {Yes} (pro3a);
\draw [arrow] (dec2) -- node[anchor=south] {No} (pro3b);
\draw [arrow] (proa) |- (middle);
\draw [arrow] (prob) -- (pro2);
\draw [arrow] (pro3b) |- (out1);
\draw [arrow] (pro3a) -- (out1);
\draw [arrow] (out1) -- (stop);
\end{tikzpicture}
\caption{Flowchart for TtoT-based PSD algorithm.}
\label{fig:TtoTflow}
\end{figure*}

\subsection{Determining discriminating threshold}
A multi-step approach is required to calculate the discriminating threshold. First, the maximum amplitude must be determined. To do so, for each record length, the local maxima of the normalised pulse is calculated using the `max' function in Python. These values are stored as the maximum amplitude of the pulses.  Second, the beginning of tail position should be determined. It is observed that the point of intersection of smoothed and discrete pulse is the start of the tail location. An array was defined to retain the intersection values for the whole set of 57934 pulses in order to store the values of the start of tail for the entire set of pulses. It was found that the smoothed and shaped pulse intersects the un-smoothed (discrete pulses) at 70\% of its maximum amplitude `on average', so the start of the tail area was chosen to be 70\% of the maximum amplitude of the un-smoothed discrete pulse \cite{Redus2009}. Finally, the tail and total area of the smoothed pulse are computed and the ratio is compared between two adjacent pulses. 
The discernible distinction in the ratio between the tail area and total area of the waveforms is utilized in the discrimination of neutrons and $\gamma$-rays through the implementation of the Tail-to-Total area (TtoT) algorithm using Python programming code. $R$ represents the ratio of tail-to-total area calculated for each measured pulses for a given acquisition time also known as window or record length. The computed feature $R$, serves as a metric for discriminating between pulses originating from neutrons and $\gamma$-rays. 
It was noted that the optimal discrimination is achieved by establishing the threshold value at 0.53, which is determined in accordance with $R$. Any value $R$ greater than 0.53 qualifies as a neutron, whereas anything less qualifies as a $\gamma$-ray. The pulses falling in on the discriminating threshold are removed as these pulses cannot be classified as either neutrons or $\gamma$-rays.  
The flowchart of our proposed TtoT-based PSD algorithm is shown in figure \ref{fig:TtoTflow}.

\begin{figure}[ht]
\centering
\subfloat[Parent Experiment]{\includegraphics[width=0.5\textwidth]{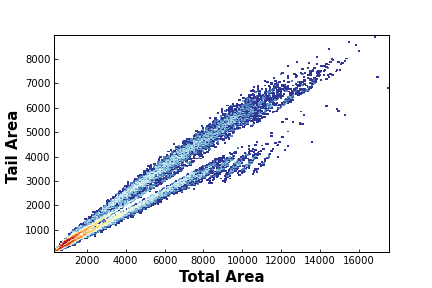}\label{fig:CI}}
\hfill
\subfloat[Test Experiment]{\includegraphics[width=0.5\textwidth]{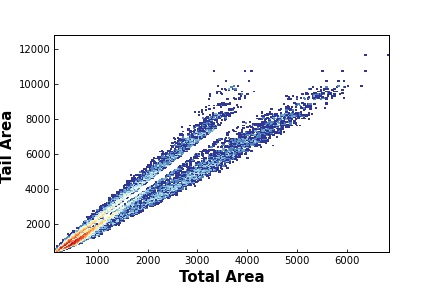}\label{fig:CI_IITR}}
\caption{ Scatter plot using Tail-to-Total Area (TtoT) method. The clusters above are neutrons and below are $\gamma$- rays.}
\label{fig:Scatter}
\end{figure}

\begin{figure}[bt]
\centering
\subfloat[Parent Experiment]{\includegraphics[width=0.45\textwidth]{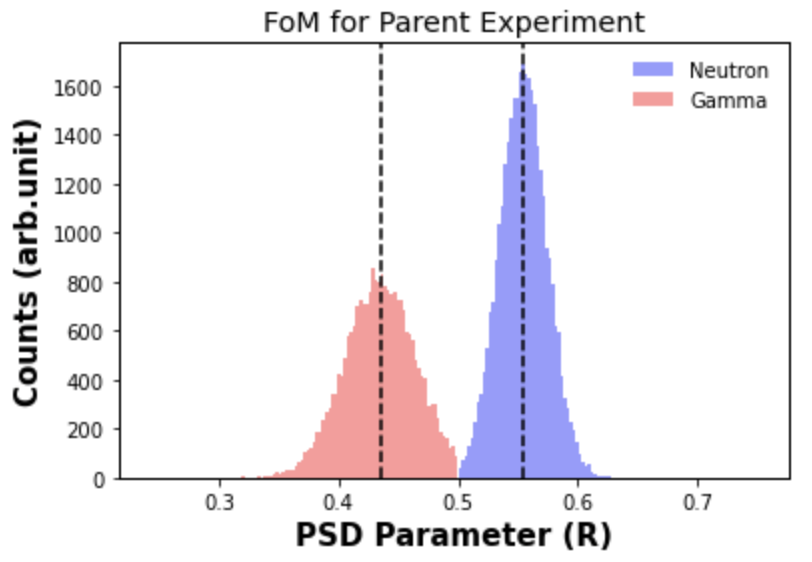}\label{fig:FOM}}
\hfill
\subfloat[Test Experiment]{\includegraphics[width=0.45\textwidth]{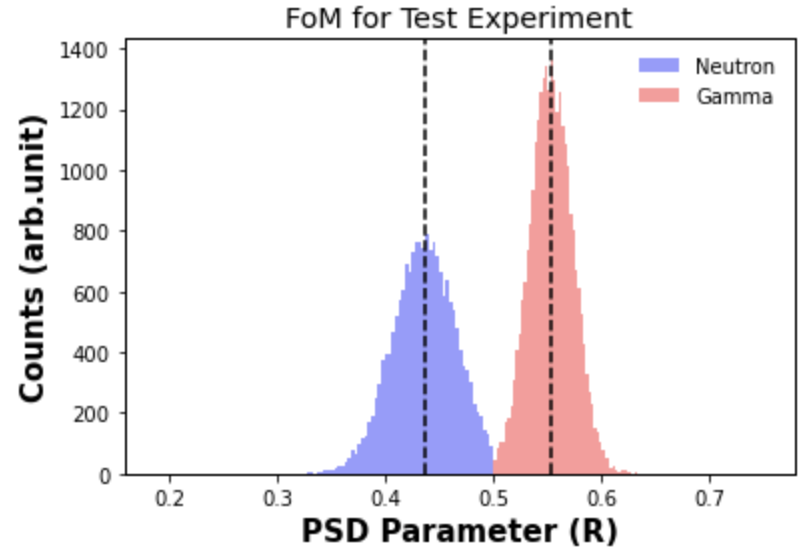}\label{fig:FOM_IITR}}
\caption{ TtoT tagged neutrons and $\gamma$- rays.}
\label{fig:FOM_FINAL}
\end{figure}

\section{Results and Discussion}
Figure \ref{fig:CI} shows a scatter plot of neutrons and $\gamma$-rays from the BC501A detector (Parent experiment) obtained using the TtoT algorithm, demonstrating good qualitative discrimination between neutrons and $\gamma$-rays. Due to the digital capabilities of our proposed technique, the output window conveniently displays the count of neutrons and $\gamma$-rays. Accordingly, the total count of neutrons and $\gamma$-rays was determined to be 33374 and 23803, respectively. Due to the limitations inherent to the experimental facility at IIT Kanpur, uncertainties arise regarding the precise accuracy of the neutron and $\gamma$-ray counts in the current methodology.

In light of this, a comparative analysis is conducted between the performance of TtoT algorithm and CI method on an experimental data (`test experiment')  which provided number of neutrons and $\gamma$-rays, described in section 2.
Figure \ref{fig:CI_IITR} shows the scatter plot representing the qualitative discrimination achieved between neutrons and $\gamma$-rays in the `test' experiment. Here, the total number of pulses considered are 46323 , with 15741 and 30055 identified as neutrons and $\gamma$-rays respectively, using our proposed TtoT method. The numbers of neutrons and $\gamma$-rays obtained from the CI method for the same are 15741 and 30852, respectively. The results obtained are summarised in table \ref{tab:Table2}. It is noteworthy to highlight that the TtoT-based PSD algorithm demonstrates a level of accuracy nearly equivalent to the built-in CI method within the digitizer for `test' experiment.

Figure-of-Merit (FoM) is used to quantify the classification performance of neutrons and $\gamma$ rays from a PSD approach. The figure of merit (FoM), given by the equation \ref{eq3}, where the `$\triangle$' depicts the spacing between centroids of the neutron and $\gamma$-ray peaks. The full-widths at half-maximum for neutrons and $\gamma$-rays are shown by $FWHM_n$ and $FWHM_{\gamma}$, respectively.

\begin{equation} \label{eq3}
    FoM = \frac{\triangle}{FWHM_{n} + FWHM_{\gamma}}
\end{equation}

\begin{table}[bt]
    \centering
    \begin{tabular}{|c|c|c|c|c|}
    \hline
    \textbf{PSD} & \textbf{Total Number} & \textbf{Number of}& \textbf{Number of} & \textbf{FoM}\\
    \textbf{Method} & \textbf{of Pulses} & \textbf{neutrons}& \textbf{ $\gamma$-rays} & \\
    \hline
         CI  & 46323 &15741 & 30852& 1.12\\
        \hline
         TtoT & 46323 &15741 & 30055& 1.13\\
    \hline 
    \end{tabular}
    \caption{PSD Output Comparison for Test experiment: CI vs. TtoT Method}
    \label{tab:Table2}
\end{table}

The discriminating capability of neutrons and $\gamma$-rays are projected on the PSD parameter axis as illustrated in figure \ref{fig:FOM_FINAL}. We consider the ratio of tail to total area, represented by R as the PSD parameter. After optimizing the threshold and excluding the 757 events falling within the cutoff range, the FoM for the `parent experiment' was determined to be 1.02, as depicted in Figure \ref{fig:FOM}. It is observed that the projected neutrons and $\gamma$-rays are very well separated. 
The Figure of Merit (FoM) for the `test experiment', after optimizing the threshold, using our proposed TtoT-based PSD method is 1.13 (Figure \ref{fig:FOM_IITR}), excluding 797 events that fell within the cutoff. The FoM for the same obtained from the CI method, is found to be 1.12. The results of `test' experiment are summarized in table\ref{tab:Table2}. It is observed that the TtoT method is able to discriminate neutrons and $\gamma$-rays with similar accuracy to that of the CI method. 

\section{Conclusion}
Our proposed TtoT algorithm demonstrates a commendable level of accuracy in discriminating between neutrons and $\gamma$-rays, comparable to the performance of the CI method implemented in the digitizer. Notably, our algorithm achieves this accuracy without relying on a dedicated hardware setup. By optimizing various parameters and employing the TtoT method, we successfully differentiate between neutrons and $\gamma$-rays based on the distribution of total area and tail area values for two different source detector digitizer setup. This outcome highlights the potential of our algorithm as an effective alternative that eliminates the need for specialized hardware with the ability to handle both neutrons and $\gamma$-ray events simultaneously, thus offering flexibility and cost-efficiency in neutron and $\gamma$-ray discrimination applications. 

 \bibliography{cas-refs}

\end{document}